%% file: cuvir.tex
\documentclass[]{aa}
\usepackage{graphics}
\usepackage{txfonts}
\begin{document}

\title{Stellar magnetosphere reconstruction from radio data}
\subtitle{Multi-frequency VLA observations and 3D-simulations of CU Virginis}

\author{
P. Leto\inst{1},
C. Trigilio\inst{2},
C.S. Buemi\inst{2},
G. Umana\inst{2},
\and
F. Leone\inst{2}
}

\offprints{P. Leto \email{p.leto@ira.inaf.it}}

\institute{INAF - Istituto di Radioastronomia 
Sezione di Noto, CP 161, Noto (SR), Italy
\and
INAF - Osservatorio Astrofisico di Catania, Via S. Sofia 78, 95123 Catania, Italy
}

\date{Received; Accepted}

\titlerunning{Stellar magnetosphere reconstruction from radio data}
\authorrunning{P. Leto, et al.}

\abstract  
{} 
{
In order to fully understand the physical processes in the
magnetospheres of the Magnetic Chemically Peculiar stars, we performed
multi-frequency radio observations of  CU~Virginis.
The radio emission of this kind of stars arises from the interaction between energetic 
electrons and magnetic field.
Our analysis is used
to test the physical scenario proposed for the radio
emission from the MCP stars
and to derive quantitative information about physical
parameters not directly observable.
}
{
The radio data were acquired with the VLA
and cover the whole rotational period of CU~Virginis.
For each observed frequency the radio light curves of the total flux density and fraction of circular polarization were fitted
using a three-dimensional MCP magnetospheric model simulating the
stellar radio emission as a function of the magnetospheric physical parameters.
}
{
The observations show a clear correlation 
between the radio emission and the orientation of the magnetosphere
of this oblique rotator.
Radio emission is explained as the result of the acceleration of the wind
particles in the current sheets just beyond the Alfv\'en radius, that eventually
return toward the star following the magnetic field and emitting radiation by gyrosyncrotron mechanism.
The accelerated electrons we probed with our simulations
have a hard energetic spectrum ($N(E)\propto E^{-2}$)
and the acceleration process has an efficiency of about $10^{-3}$.
The Alfv\'en radius we determined is in the range of $12-17\,R_\ast$ and, 
for a dipolar field of 3000 Gauss at the magnetic pole of the star,
we determine a mass loss from the star of about
$10^{-12}$ M$_{\sun}$ yr$^{-1}$. In the inner magnetosphere, inside the Alfv\'en radius,
the confined stellar wind accumulates {and reaches temperatures in the range of}
$10^5-10^6$ K, and a detectable X-ray emission is expected.
}
{} 


     \maketitle
%

\section{Introduction}
Magnetic Chemically Peculiar (MCP) stars 
are characterized by periodic variability of the effective
magnetic field through the stellar rotational period.
The magnetic field 
topology of this kind of star
has been commonly regarded as a magnetic dipole
tilted with respect to the rotation axis (Babcock \cite{b49}; Stibbs \cite{s50}),
even if MCP stars with a multipolar magnetic field are also known
(Landstreet \cite{l90}; Mathys \cite{m91}).

In some cases, MCP stars are characterized by an anisotropic stellar wind
(Shore et al. \cite{s_etal87}; Shore \& Brown \cite{sb90}),
as a consequence of the wind interaction with 
the dipolar magnetic field (Shore \cite{s87}).
The resulting magnetosphere is structured
in a ``wind zone'' at high magnetic latitudes,
where  the gas flows draw the magnetic field out to open
structures, and a ``dead zone'', where the magnetic field
topology is closed and the gas trapped.

\begin{table*}
\caption[ ]{Results of the observations. For each observing scan, 
the mean UT time, the total flux density (Stokes\,$I$)
at the three observed frequency ($I_5$, $I_{8.4}$ and $I_{15}$), the corresponding errors ($\sigma$)
and the circular polarized flux density, Stokes\,$V$ ($V_5$, $V_{8.4}$ and $V_{15}$), are reported:} 
\label{tab1}
\input{tab5_cuvir.tex}
\end{table*}

Currently, non-thermal radio emission is observed from about
25\% of MCP stars
and the detection rate seems to be directly correlated to
the effective stellar temperature
(Drake et al. \cite{d_etal87}; Linsky et al. \cite{l_etal92}; Leone et al. \cite{l_etal94}).
Since only MCP stars with high photospheric temperature develop
a radiatively-driven stellar wind, this result
reveals the importance of a stellar wind in driving the radio emission.
To explain the origin of the radio emission from
pre-main sequence magnetic B stars and from MCP stars,
Andr\'e et al. (\cite{a_etal88}) proposed a model characterized by
the interaction between the wind and the magnetic field.
The radio emission is understood
in terms of gyrosynchrotron emission from mildly relativistic electrons,
which are accelerated in ``current sheets'' formed where the gas flow
brakes the magnetic field lines, close to the magnetic
equator plane, and propagate along the magnetic field lines back to
the inner magnetospheric regions  (Havnes \& Goertz \cite{hg84}; Usov \& Melrose \cite{um92}).

According to the oblique dipole model, Leone (\cite{l91}) suggested that the radio
emission of MCP stars should be also periodically variable.
The confirmation of this hypothesis comes from
the observations of HD37017 and HD37479
(Leone \& Umana \cite{lu93}) and of  HD133880
(Lim et al. \cite{l_etal96}), performed at 5 GHz, that
revealed the presence of rotational flux modulation.
In particular, the observed flux light curves of HD37017
and HD37479, which are characterized
by a simple dipolar magnetic field,
show a radio emission minimum
coincident with the zero of the longitudinal
magnetic field, whereas the emission is
maximum at the magnetic field extrema.

Among radio MCP stars, only CU Virginis (HD124224) is
characterized by a strong 1.4 GHz flux enhancement around phases 0.4 and 0.8,
with a right hand circular polarization of almost 100\% (Trigilio et al. \cite{t_etal00}, 
TLL00). This discovery has been explained in terms of Electron
Cyclotron Maser Emission (ECME) from energetic electrons accelerated in the current sheets
near the Alfv\'en surface (where the thermal plus  kinetic plasma pressure equals the
magnetic one) and reflected outward by magnetic mirroring  in the inner magnetospheric
layers.

In a previous paper
(Trigilio et al. \cite{t_etal04}, TLU04),
we developed a three-dimensional numerical
model of the radio emission from MCP stars with a dipolar field.
On the basis of successfull simulated 5 GHz flux light curves of HD37017 and HD37479,
whose study allowed us to confirm the qualitative scenario proposed 
to explain the origin of non-thermal radio emitting electrons. 
We showed that the combined effect
of a tilted dipolar magnetic topology and the
presence of a thermal plasma trapped in the ``dead-zone''
can produce the observed phase modulation
of the stellar radio emission.

In this paper we present the multifrequency 
microwave observations of CU Virginis, a well
studied  Si-group MCP star.
\begin{figure}
\resizebox{13cm}{!}{\includegraphics{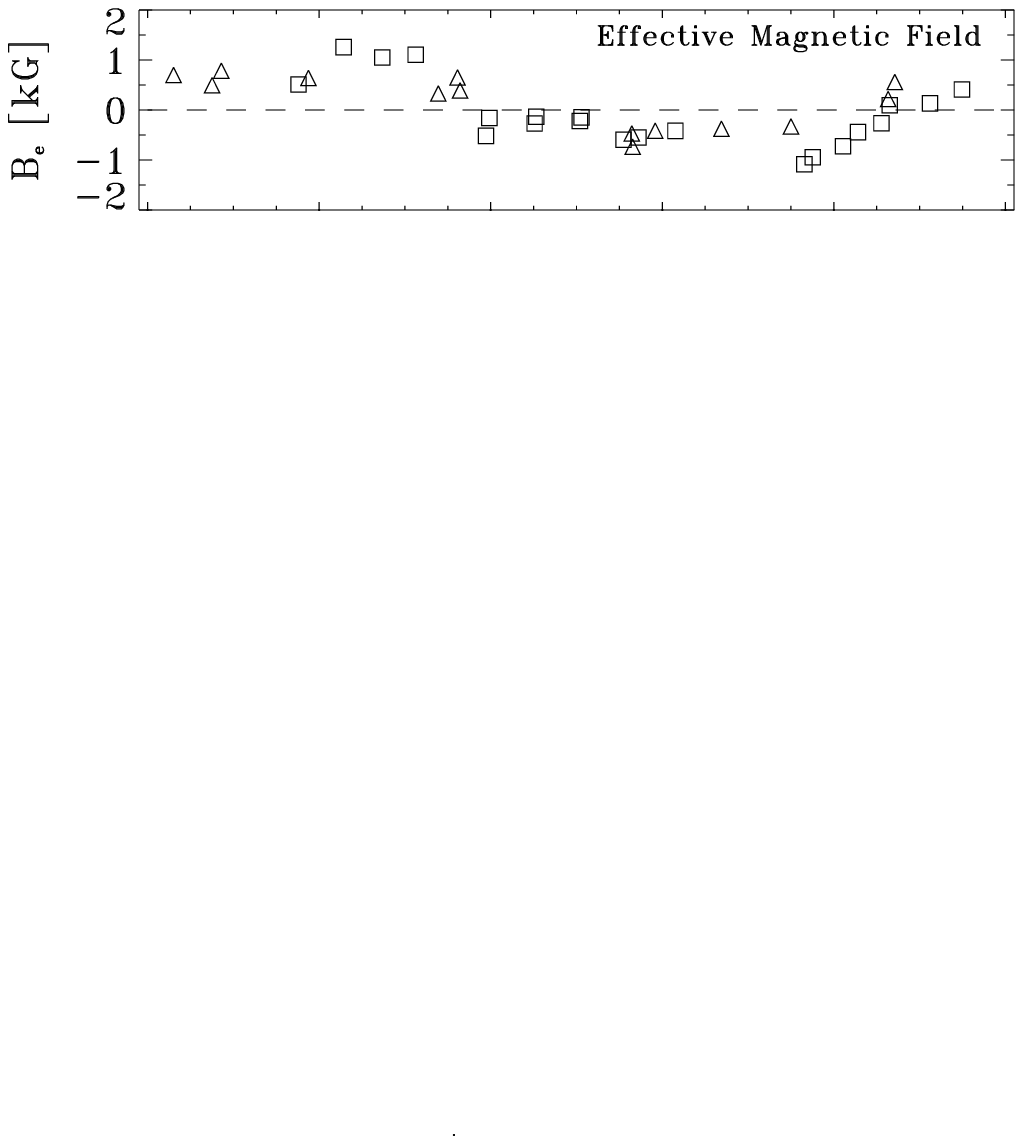}}
\resizebox{13cm}{!}{\includegraphics{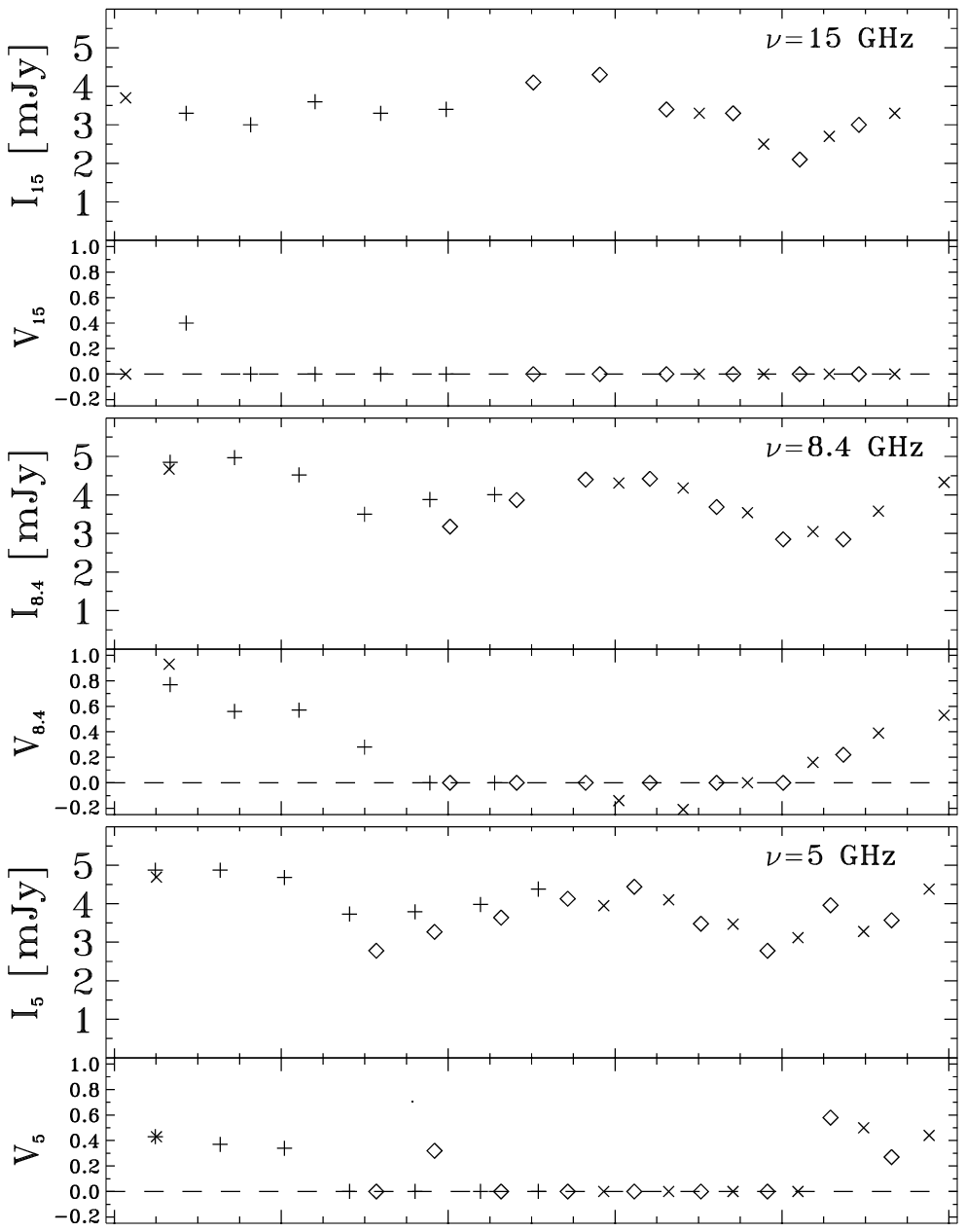}}
\resizebox{13cm}{!}{\includegraphics{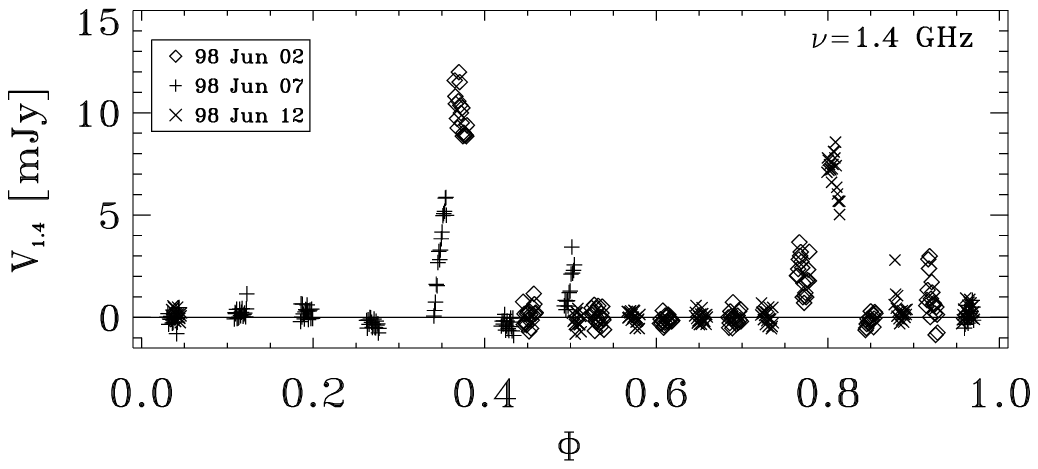}}
 \caption[ ]{Top panel: effective magnetic field of
CU Vir versus rotational phase
(triangles: data by Borra \& Landstreet \cite{bl80}; 
squares: data by Pyper et al. \cite{p_etal98}).
Other panels:  phase modulation of Stokes\,$I$ and Stokes\,$V$ flux density
at 15, 8.4 and 5 GHz. For comparison, the 1.4 GHz Stokes\,$V$ (lowest panel) 
by TLL00 is reported. The three days of observations are distinguished
with different symbols to point out the repeatability of the radio emission in
time.}
\label{dat_dat}
\end{figure}
To investigate the physical conditions of the stellar magnetosphere,
the radio light curves of CU Vir are compared
with the results of the numerical simulations performed
using our 3D model.
We present here an implementation of our code to 
also treat circular polarization.

\section{Observations and data reduction}
The observations were carried out using
the VLA\footnote{The Very Large Array is a facility of the National Radio
Astronomy Observatory which is operated by Associated Universities, Inc.
under cooperative agreement with the National Science Foundation.}
in three different runs, on June 02, 07 and 12, 1998. Each run was
approximately eight hours long.
We observed at four frequencies (1.4, 5, 8.4 and 15 GHz) using two independent
and contiguous bands of 50 MHz in Right and Left
Circular Polarizations (RCP and LCP). The observations were 
performed using the entire array (configuration array moving A $\rightarrow$ B)
alternating all the frequencies.

Because of the $\approx 0.52^d$ rotational period and with the aim to
obtain a complete rotational phase coverage, we assumed
at any frequency an observing cycle consisting of $\approx$10 minutes
on source, preceded and followed by $\approx$2 minutes on the
phase calibrator ($1354-021$). To avoid redundancy in the phase
coverage for each frequency, the observing sequence was
changed in each run. 

The flux density scale was determined by daily observations of 3C286. 
The data were calibrated and mapped using the standard procedures
of the Astronomical Image Processing System (AIPS).
The flux density and fraction of circular polarization
were obtained by fitting a two-dimensional Gaussian, task JMFIT,
at the source position in the 
cleaned maps (separately for the Stokes\,$I$ and Stokes\,$V$ 
\footnote{VLA measurements of wave polarization state are in agreement with
the IAU and IEEE orientation/sign convention: positive V
corresponds to right circular polarization.}) integrated over
the duration of each individual scan. As the uncertainty in the flux density
measurements we assume the r.m.s. of the map.
Results are summarized in Table~\ref{tab1}.
The measurements performed at 1.4 GHz have been
presented elsewhere (TLL00).

\section{The radio behavior of CU~Vir}
\label{r_c}
In Fig.~\ref{dat_dat}, for each observed frequency, we report the radio flux density
measured in both Stokes $I$ and $V$ versus the
rotational phase. This was computed adopting the ephemeris referred to light minimum
reported by Pyper et al. (\cite{p_etal98}):  
\\

\noindent
$\mathrm {HJD}=2\,435\,178.6417+\left\{
        \begin{array}{lr}
        0^{\mathrm{d}}.52067780 E & \mbox{JD $<$ 2\,446\,000} \\
        0^{\mathrm{d}}.52070308 E & \mbox{JD $>$ 2\,446\,000}
        \end{array}
\right.
$\\

In the lowest panel of Fig.~\ref{dat_dat}
the measures performed in Stokes\,$V$ at 1.4 GHz from TLL00 are also displayed.

The top panel of Fig.~\ref{dat_dat} shows the average longitudinal component of the magnetic field
as a function of the rotational phase (data are from Borra \& Landstreet \cite{bl80}
and Pyper et al. \cite{p_etal98}).

The flux variability 
with the stellar rotational period suggests that the
radio emission at 15, 8.4 and 5 GHz
arises from a stable co-rotating magnetosphere
and that the radio source is optically thick.
For Stokes $I$ the radio emission minimum appears to be
coincident with the null magnetic field ($\Phi \approx 0.4$ and $\Phi \approx 0.8$),
whereas the two maxima, clearly visible at 5 and 8.4 GHz, are
related to stellar orientations coinciding with the two extrema of the stellar magnetic field
($\Phi \approx 0.1$ and $\Phi \approx 0.6$).
Stokes\,$V$ is detected at all the observed frequencies close to
$\Phi \approx 0.1$, whereas at $\Phi \approx 0.6$ the
circular polarization is detected only at 8.4 GHz.
The circularly polarized
flux density reaches its positive (RCP) maximum value
when the effective magnetic field strength is also maximum.
This means that radio flux emitted in a direction close to the magnetic north pole,
where magnetic field lines are mainly radially oriented (relative to the outside stellar surface),
is partially right hand polarized.
Negative values of Stokes\,$V$ (LCP) have been detected only at 8.4 GHz around $\Phi \approx 0.6$,
when the magnetic south pole is closest to the line of sight.

The phase modulation of the Stokes\,$I$ flux density has been observed
in other MCP stars at 5 GHz (Leone \cite{l91}; Leone \& Umana \cite{lu93};
Lim et al. \cite{l_etal96}). Our dataset represents an extension of previous studies
for the (almost simultaneous) multi-frequency observation of both total
and circular polarized flux with a detailed phase coverage of the rotational period.
This allows us to monitor the variations of an MCP star
radio spectrum
through its entire stellar rotation.

On average the observed radio spectra
are flat with a
slightly 
negative power law 
index $\alpha \approx -0.1$ ($I_{\nu} \propto \nu ^{\alpha}$).
This, together with the low fraction of circular polarization (about 10\%), supports
the idea that non-thermal gyrosynchrotron
from a non-homogeneous source is responsible for the observed radio emission.

\begin{figure}
\resizebox{10cm}{!}{\includegraphics{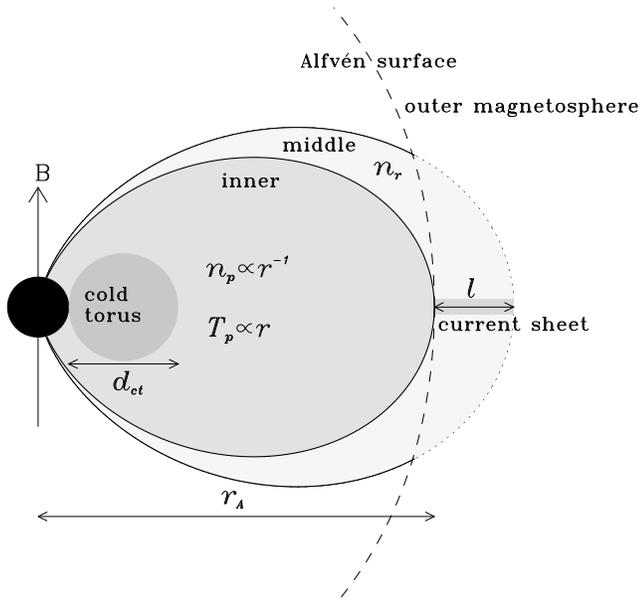}}
 \caption[ ]{Schematic view of the magnetosphere of an MCP star assuming a
 central dipolar magnetic field. The dashed line shows the Alfv\'en surface:
 the largest magnetic field line delimiting the middle magnetosphere is drawn
 with a dotted line outside the  Alfv\'en surface.
The length $l$ represents the current sheets region where the non-thermal electrons are
efficiently accelerated. The inner magnetosphere is filled 
with hot thermal plasma and cold material accumulates in the form of a torus near the stellar surface.
}
\label{sezione}
\end{figure}

\section{The model}

Our 3-D model samples the magnetosphere of the star in a three dimensional grid, 
assumes a set of physical parameters, computes emission and absorption coefficients
for gyrosynchrotron emission, and
solves the equation of transfer for polarized radiation in terms of Stokes $I$ and $V$ (appendix~\ref{app}).
From the best fit 
of the observed fluxes, we
determine the most probable physics of the magnetosphere.

\subsection{Basic picture}
The model was developed (TLU04) on the basis of the physical
scenario proposed by Andr\'e et al. (\cite{a_etal88}) and  Linsky et al. (\cite{l_etal92}). 
The model involves the interaction
between the dipolar magnetic field and the stellar wind
resulting in the formation of ``current sheets'' outside the
equatorial region of the Alfv\'en surface (Havnes \& Goertz \cite{hg84}), where the 
wind particles
can be accelerated up to relativistic energy (Usov \& Melrose \cite{um92}),
radiating at radio wavelengths by the gyrosynchrotron mechanism.

\begin{table}
\caption[ ]{Adopted stellar parameters for CU Virginis.}
\label{par_star}
\footnotesize
\begin{tabular}[]{lcl}
\hline
\hline
Distance [pc]                           &$D$                            & 80$^{\dag}$\\
Stellar Radius [R$_{\sun}$]             &$R_{\ast}$                     & 2.2$^{\mathrm {a}}$\\
Polar Magnetic Field [Gauss]            &$B_\mathrm{p}$                 & 3000$^{\mathrm {d}}$\\
Rotation Axis Inclination [degree]      &$i$                            & ${43}^{\mathrm {d}}$\\
Magnetic Axis Obliquity [degree]        &$\beta$                        & ${74}^{\mathrm {d}}$\\
\hline
\end{tabular}
\begin{list}{}{}
\item[$^{\dag}$] ESA (\cite{esa97})
\item[$^{\mathrm {a}}$] North (\cite{n98})
\item[$^{\mathrm {b}}$] Kuschnig et al. (\cite{k_etal99})
\item[$^{\mathrm {c}}$] Pyper et al. (\cite{p_etal98})
\item[$^{\mathrm {d}}$] Trigilio et al. (\cite{t_etal00})
\end{list}
\end{table}

\begin{table*}
\begin{center}
\caption[ ]{Input parameters of the 3-D model}
\label{mod_par}
\begin{tabular}{lccl}
\hline
\hline
   &Symbol         &Range          &Simulation step\\
\hline
\multicolumn{4}{l}{\bf Free parameters - step 1}\\
\hline
Size of Alfv\'en radius [R$_{\ast}$]                            &$r_\mathrm{A}$         &$6$ -- $30$    &$\Delta r_\mathrm{A}=1$ \\
Thickness of magnetic shell [R$_{\ast}$]        &$l$            &$0.3$ -- $30$   &$\Delta \log l=0.05$ \\
Relativistic electron density [cm$^{-3}$]                       &$n_\mathrm{r}$ &1 -- $10^{7}$  &$\Delta \log n_\mathrm{r}=0.25$  \\
Relativistic electron energy power-law index             &$\delta$       &2 -- 3      &$\Delta \delta=0.5$ \\
Temperature of the trapped thermal plasma at stellar surface [K]  &$T_\mathrm{p_0}$  &$10^4$ -- $10^7$  &$\Delta \log T_\mathrm{p_0}=1$\\
Electron density of the thermal plasma at stellar surface [cm$^{-3}$]  &$n_\mathrm{p_0}$ &$10^{8}$ -- $10^{11}$  &$\Delta \log n_\mathrm{p_0}=0.225$ \\
\hline
\multicolumn{4}{l}{\bf Free parameters - step 2}\\
\hline
Cross-sectional size of the Cold Torus [R$_{\ast}$]                             &$d_\mathrm{ct}$                &$1$ -- $6$      &$\Delta d_\mathrm{ct}=1$\\
Temperature of the plasma of the Cold Torus [K] &$T_\mathrm{ct}$  &$10^3$ -- $10^5$      &$\Delta \log T_\mathrm{ct}=0.25$\\
Electron density number of the plasma in the Cold Torus [cm$^{-3}$]             &$n_\mathrm{ct}$        &$10^{9}$ -- $10^{12}$      &$\Delta \log n_\mathrm{ct}=0.5$\\
\hline
\end{tabular}
\end{center}
\end{table*}

The adopted geometry is shown in Fig.~\ref{sezione}.
The layer where relativistic
electrons propagate is delimited by two magnetic lines: the first is
the largest closed line at a tangent to the Alfv\'en surface in the plane of the
magnetic equator and the latter is defined by the length of the current sheets
where electrons are efficiently accelerated.
In each point of this shell, the ``middle" magnetosphere in analogy
with Andr\'e et al. (\cite{a_etal88}), it is assumed that
emitting electrons have a power law energy spectrum,
isotropic pitch angle distribution and 
homogeneous spatial distribution due to magnetic mirroring.

Following the magnetically confined wind shock (MCWS) model by
Babel \& Montmerle (\cite{bm97}),
the ``inner" magnetosphere (``dead zone") is filled by a trapped thermal 
plasma heated to a temperature higher then the photospheric value 
($T_\mathrm{phot}=12500$ K, Kuschnig et al. \cite{k_etal99}).
This hot thermal plasma progressively pushes the shock front to high magnetic latitudes,
until its thermal pressure equals the wind ram pressure.

The net effect of this mechanism is the filling of the inner magnetosphere with hot
thermal plasma. If the magnetosphere rotates, the density decreases linearly outward,
while the temperature increases linearly (Babel \& Montmerle, \cite{bm97}).

To balance the continuous supply of hot matter from the polar regions,
a cool plasma down-flow is expected from the inner magnetosphere toward the
stellar surface. This process causes the accumulation of
cold dense material, which has been assumed for simplicity as a
homogeneous torus with a circular cross-section surrounding the star
(Babel \& Montmerle \cite{bm97}; Smith \& Groote \cite{sg01}).

\subsection{The parameters}
In Table~\ref{par_star} we list the assumed stellar parameters.
The dependence of our simulations on 
the polar magnetic field $B_\mathrm{p}$, inclination $i$ and obliquity $\beta$
that are not known with great accuracy
will be discussed later in Sect.~\ref{dep_stellar_param}.

The free parameters of the model are:
\begin{itemize}
\item[$r_\mathrm{A}$] average radius of the Alfv\'en surface in R$_{\ast}$ units;
\item[$l$] thickness of the middle-magnetosphere at $r_\mathrm{A}$ in R$_{\ast}$ units;
\item[$n_\mathrm{r}$] total number density of non-thermal electrons;
\item[$\delta$] spectral index of non-thermal electron
energy distribution;
\item[$T_\mathrm{p_0}$] temperature, at the stellar surface, of the thermal plasma
trapped in the inner-magnetosphere;
\item[$n_\mathrm{p_0}$] electron number density, at the stellar surface, of thermal plasma;
\item[$d_\mathrm{ct}$] cross-sectional diameter of the cold torus surrounding the star;
\item[$T_\mathrm{ct}$] temperature of the cold torus;
\item[$n_\mathrm{ct}$] electron number density of the cold torus.
\end{itemize}

The range of values and the assumed simulation steps
of these free parameters
are reported in Table~\ref{mod_par}.

In TLU04 the Alfv\'en surface has been located numerically by solving
the equation of balance between kinetic and magnetic energy density. In this work
$r_\mathrm{A}$ is assumed as a free parameter, since the wind parameters of CU Vir are 
still unknown. The mass loss rate $\dot{M}$, assumed flowing from the whole stellar surface,
and the plasma density of the wind at the Alfv\'en surface
are thus derived as output parameters for a given $r_\mathrm{A}$ 
and wind velocity $v_{\infty}$.

For a detailed discussion of the computational method
we refer to TLU04.

\subsection{Numerical simulations}

Since the computation is quite time consuming,
we split the analysis in two steps. 

\subsubsection{Step 1: fitting the 8.4 GHz light curve}

In this first step we can neglect the presence of the cold torus.
As pointed out by TLU04, it does not have an appreciable effect
in the low frequency radio emission (5 and 8.4 GHz) that
arises from external magnetospheric regions, and we exclude a priori 
its effect in the simulations at 8.4 GHz, 
reducing the number of free parameters (Table~\ref{mod_par} - step 1).

For the choice of the sampling length of the three dimensional grid
we adopt the same criteria as in TLU04, with a narrow
spacing (0.08 R$_{\ast}$) for the distance from the star center
of less than 2.3 R$_{\ast}$, middle spacing (0.3 R$_{\ast}$)
between 2.3 and 7 R$_{\ast}$, and rough spacing (1 R$_{\ast}$) 
greater than 7 R$_{\ast}$.

%

We compute the radio emission for the Stokes $I$ and $V$
at the phases that characterize the observed light curve at 8.4 GHz
(see Sect.~\ref{r_c}). The match between simulations and observations
defines the possible configurations of the magnetosphere.


\subsubsection{Step 2: fitting the spectra}
For any set of parameters modeling the observed Stokes\,$I$ and $V$ at 8.4 GHz
we compute the radio emission at 5 and 15 GHz
to search for the ones fitting the whole observed spectra and their 
polarization degree.

However we cannot 
rule out the importance of the cold torus on the
stellar emission at the frequency of 15 GHz.
The free parameters that characterize the cold torus and
the adopted ranges are reported in Table~\ref{mod_par} - step 2.

 \begin{table}
 \begin{center}
 \caption[ ]{Solutions of the 3-D model:
 the free parameters able to reproduce the observations.
 }
 \label{sol_par}
 \begin{tabular}[]{llcl}
 \hline
 \hline
Symbols       &  & Solutions  & Units  \\
\hline
 $r_\mathrm{A}$     &          &     12 -- 17 &  [R$_{\ast}$]    \\
 $\delta$           &          &     2        &  ~~~             \\
 $d_\mathrm{ct}$    &          &     5        &  [R$_{\ast}$]    \\
 $l/r_\mathrm{A}$   &          & 0.05 -- 0.2  &~~~               \\
 $n_\mathrm{r}$     &$\propto {r_\mathrm{A}}/{l}$ & 0.88 -- 3.54   &[$\times 10^3$ cm$^{-3}$]\\
 \vspace{-.3cm}\\
 $n_\mathrm{p_0}$   &$=10^{5.95}T_\mathrm{p_0}^{0.68}$    &           &\\
 \vspace{-.3cm}\\
 $n_\mathrm{ct}$    &$\geq 10^{5.95}T_\mathrm{ct}^{0.68}$ &           &\\
 \vspace{-.3cm}\\
 \hline
 \end{tabular}
 \end{center}
 \end{table}

\begin{figure*}
\resizebox{17cm}{!}{\includegraphics{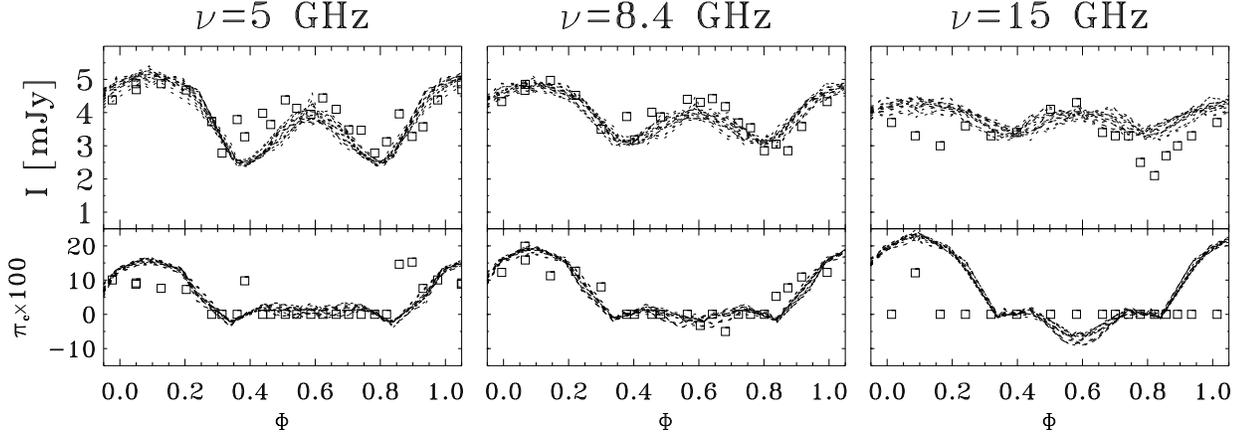}}
 \caption[ ]{Comparison between CU~Vir observed and computed
flux density and degree of circular polarization, as a function of
rotational phase.}
\label{oss-tot}
\end{figure*}

\section{Results}

\subsection{Prediction of the simulations}
Comparison between the observations and simulations 
allow us to find the combinations of the model free parameters that
can account for the observational behavior.
The convergence of these quantities gives important informations about the physics
of the external regions of this star.

The solutions derived from our analysis are reported in Table~\ref{sol_par};
the corresponding observed and computed
flux light curves are plotted in Fig.~\ref{oss-tot}.

The free parameters $\delta$ and $d_\mathrm{ct}$ are determined with an uncertainty smaller than the simulation steps. We find acceptable solutions for
$r_\mathrm{A}$ in the range 12 -- 17 R$_{\ast}$,
$l$ in 0.05 -- 0.2 $r_\mathrm{A}$ and $n_\mathrm{r}$ in 0.88$\times 10^3$ -- 3.54$\times 10^3$ cm$^{-3}$.
We note that the relation $n_\mathrm{r} \propto r_\mathrm{A}/l$ holds (Table~\ref{sol_par}).
An important conclusion can be drawn:
the column density of the relativistic electrons at the Alfv\'en radius
is univocally determined by $r_\mathrm{A}$.
For the different values of the Alfv\'en radius
deduced from our simulations the column density is in the 
range $3.2\times 10^{14} - 4.6\times 10^{14}$ cm$^{-2}$.

%

Regarding the physical properties of the thermal plasma trapped in the inner magnetosphere, we found that
the values of the electron number density and temperature that account for the
observed spectral index are related as
\begin{equation}
n_\mathrm{p_0}=10 ^{5.95}~ T_\mathrm{p_0} ^{0.68},
\label{fit}
\end{equation}
while for the plasma in the cold torus
\begin{displaymath}
n_\mathrm{ct} \geq 10^{5.95}\, T_\mathrm{ct}^{0.68}.
\end{displaymath}
In terms of free-free
absorption coefficients (Dulk \cite{d85}),
\\

\noindent
$
\kappa_{\nu} = 9.78 ~ 10^{-3} \frac{n^2}{\nu ^2 T^{\frac{3}{2}}} \times\left\{
\begin{array}{lr}
18.2 +\frac{3}{2}\ln T - \ln \nu &\mbox{\footnotesize ($T < 2 ~ 10^5$ K)}\\
24.5 +\ln T - \ln \nu            &\mbox{\footnotesize ($T > 2 ~ 10^5$ K)}.
\end{array}
\right.
$
\\

\noindent
The above relations imply that the plasma surrounding the star is almost opaque
at $\nu=$15 GHz, and that the cold torus around the star is completely opaque at
the same frequency.


~\\
The Alfv\'en radius $r_\mathrm{A}$ is a free parameter of our model.
It is defined as the point where the magnetic energy density is equal to
the kinetic energy density. If the flowing wind has terminal velocity $v_{\infty}$
and density $\rho_\mathrm{w}=m_\mathrm{H} n_\mathrm{w}$,
with proton mass $m_\mathrm{H}$, then:

\begin{displaymath}
\frac {B^2}{8\pi} = \frac{1}{2}\rho_\mathrm{w} v_{\infty}^2.
\end{displaymath}

For a given $r_\mathrm{A}$, we obtain the corresponding magnetic field strength
$B(r_\mathrm{A})$, and consequently the thermal plasma density of the wind
$n_\mathrm{w}(r_\mathrm{A})$ at the Alfv\'en surface.
In Table~\ref{massloss} we list the resulting values for 
$n_\mathrm{w}(r_\mathrm{A})$ as a function of 
$r_\mathrm{A}$ assuming three different values
of the wind velocity ($v_{\infty}=400$, 600 and 800 km s$^{-1}$).

The mass loss rate for a spherically symmetric wind is easily computed 
from the continuity equation
\begin{displaymath}
\dot{M}=4\pi r_\mathrm{A}^2 v_{\infty} m_\mathrm{H} n_\mathrm{w}.
\end{displaymath}
In the presence of a magnetic dipole,
only the wind flowing out from the polar caps can escape from the star and
the actual mass loss rate $\dot{M}_\mathrm{act}$ is a fraction of $\dot{M}$,
defined as the ratio of the area of the polar caps to the whole
stellar surface (see TLU04).
The results for $\dot{M}$ and $\dot{M}_\mathrm{act}$ are shown in
Table~\ref{massloss}.


 \begin{table}
 \begin{center}
 \caption[ ]{Results for wind, mass loss and inner magnetosphere:
 as a function of the Alfv\'en radius $r_\mathrm{A}$,
 the spherical wind mass loss rate $\dot{M}$, the effective mass loss
 $\dot{M}_\mathrm{act}$ and the number density of the electrons
 of the wind at the Alfv\'en point $n_\mathrm{w}(r_\mathrm{A})$ are reported.
 In the last two columns, temperature ($T_\mathrm{p_0}$) and number density ($n_\mathrm{p_0}$)
at the base of the magnetosphere are also reported.
 }
 \label{massloss}
 \footnotesize
 \begin{tabular}[]{ccccccccccc}
 \hline
 \hline
 \vspace{-.3cm}\\
 \multicolumn{9}{l}{{\bf a)} $v_{\infty}=400$ km s$^{-1}$}\\
 \hline
 \vspace{-.3cm}\\
 $r_\mathrm{A}$  &\multicolumn{2}{c}{$\dot{M}$}
 &\multicolumn{2}{c}{$\dot{M}_\mathrm{act}$} &\multicolumn{2}{c}
 {$n_\mathrm{w}(r_\mathrm{A})$}&\multicolumn{2}{c}{$T_\mathrm{p_0}$} &
 \multicolumn{2}{c}{$n_\mathrm{p_0}$}\\
 {\scriptsize [R$_{\ast}$]} &\multicolumn{2}{c}{[\scriptsize M$_{\sun}$
 yr$^{-1}$]}  &\multicolumn{2}{c}{[\scriptsize M$_{\sun}$ yr$^{-1}$]}
 &\multicolumn{2}{c}{\scriptsize [$10^6$ cm$^{-3}$]}
 &\multicolumn{2}{c}{\scriptsize [K]}& \multicolumn{2}{c}{\scriptsize
 [cm$^{-3}$]}\\
 \hline
 {12} &\multicolumn{2}{c}{$1.6 ~ 10^{-10}$} &\multicolumn{2}{c}{$6.8 ~
 10^{-12}$} &\multicolumn{2}{c}{3.90} & \multicolumn{2}{c}{$4.94 ~ 10^4$}
 &\multicolumn{2}{c}{$1.40 ~ 10^9$}\\
 {13} &\multicolumn{2}{c}{$1.0 ~ 10^{-10}$} &\multicolumn{2}{c}{$3.9 ~
 10^{-12}$} &\multicolumn{2}{c}{2.09} &\multicolumn{2}{c}{$3.39 ~
 10^4$}&\multicolumn{2}{c}{$1.08 ~ 10^9$}  \\
 {14} &\multicolumn{2}{c}{$6.4 ~ 10^{-11}$} &\multicolumn{2}{c}{$2.3 ~
 10^{-12}$} &\multicolumn{2}{c}{1.17} &\multicolumn{2}{c}{$2.37 ~ 10^4$}
 &\multicolumn{2}{c}{$8.54 ~ 10^8$} \\
 {15} &\multicolumn{2}{c}{$4.3 ~ 10^{-11}$} &\multicolumn{2}{c}{$1.4 ~
 10^{-12}$} &\multicolumn{2}{c}{0.68} &\multicolumn{2}{c}{$1.72 ~ 10^4$
 }&\multicolumn{2}{c}{$6.89 ~ 10^8$}  \\
 {16} &\multicolumn{2}{c}{$2.9 ~ 10^{-11}$} &\multicolumn{2}{c}{$9.2 ~
 10^{-13}$} &\multicolumn{2}{c}{0.41} &\multicolumn{2}{c}{$1.26 ~ 10^4$}
 &\multicolumn{2}{c}{$5.58 ~ 10^8$}  \\
 {17} &\multicolumn{2}{c}{$2.0 ~ 10^{-11}$} &\multicolumn{2}{c}{$6.0 ~
 10^{-13}$} &\multicolumn{2}{c}{0.25} &\multicolumn{2}{c}{$0.94 ~ 10^4$}
 &\multicolumn{2}{c}{$4.58 ~ 10^8$}  \\
 \hline
 \vspace{-.3cm}\\
 \multicolumn{7}{l}{{\bf b)} $v_{\infty}=600$ km s$^{-1}$}\\
 \hline
 \vspace{-.3cm}\\
 $r_\mathrm{A}$  &\multicolumn{2}{c}{$\dot{M}$}
 &\multicolumn{2}{c}{$\dot{M}_\mathrm{act}$} &\multicolumn{2}{c}
 {$n_\mathrm{w}(r_\mathrm{A})$}&\multicolumn{2}{c}{$T_\mathrm{p_0}$} &
 \multicolumn{2}{c}{$n_\mathrm{p_0}$}\\
 {\scriptsize [R$_{\ast}$]} &\multicolumn{2}{c}{[\scriptsize M$_{\sun}$
 yr$^{-1}$]}  &\multicolumn{2}{c}{[\scriptsize M$_{\sun}$ yr$^{-1}$]}
 &\multicolumn{2}{c}{\scriptsize [$10^6$
 cm$^{-3}$]}&\multicolumn{2}{c}{\scriptsize [K]}&
 \multicolumn{2}{c}{\scriptsize [cm$^{-3}$]}\\
 \hline
 {12} &\multicolumn{2}{c}{$2.2 ~ 10^{-10}$}  &\multicolumn{2}{c}{$9.4 ~
 10^{-12}$} &\multicolumn{2}{c}{3.70}&\multicolumn{2}{c}{$7.62 ~ 10^4$}
 &\multicolumn{2}{c}{$1.87 ~ 10^9$}   \\
 {13} &\multicolumn{2}{c}{$1.4 ~ 10^{-10}$}  &\multicolumn{2}{c}{$5.5 ~
 10^{-12}$} &\multicolumn{2}{c}{2.04} &\multicolumn{2}{c}{$5.28 ~ 10^4$}
 &\multicolumn{2}{c}{$1.46 ~ 10^9$}  \\
 {14} &\multicolumn{2}{c}{$9.5 ~ 10^{-11}$}  &\multicolumn{2}{c}{$3.4 ~
 10^{-12}$} &\multicolumn{2}{c}{1.17} &\multicolumn{2}{c}{$3.83 ~ 10^4$}
 &\multicolumn{2}{c}{$1.18 ~ 10^9$}  \\
 {15} &\multicolumn{2}{c}{$6.4 ~ 10^{-11}$}  &\multicolumn{2}{c}{$2.2 ~
 10^{-12}$} &\multicolumn{2}{c}{0.69} &\multicolumn{2}{c}{$2.78 ~ 10^4$}
 &\multicolumn{2}{c}{$9.51 ~ 10^8$}  \\
 {16} &\multicolumn{2}{c}{$4.5 ~ 10^{-11}$}  &\multicolumn{2}{c}{$1.4 ~
 10^{-12}$} &\multicolumn{2}{c}{0.42} &\multicolumn{2}{c}{$2.09 ~ 10^4$}
 &\multicolumn{2}{c}{$7.84 ~ 10^8$}  \\
 {17} &\multicolumn{2}{c}{$3.2 ~ 10^{-11}$}  &\multicolumn{2}{c}{$9.6 ~
 10^{-13}$} &\multicolumn{2}{c}{0.27} &\multicolumn{2}{c}{$1.58 ~ 10^4$}
 &\multicolumn{2}{c}{$6.51 ~ 10^8$}  \\
 \hline
 \vspace{-.3cm}\\
 \multicolumn{7}{l}{{\bf c)} $v_{\infty}=800$ km s$^{-1}$}\\
 \hline
 \vspace{-.3cm}\\
 $r_\mathrm{A}$  &\multicolumn{2}{c}{$\dot{M}$}
 &\multicolumn{2}{c}{$\dot{M}_\mathrm{act}$} &\multicolumn{2}{c}
 {$n_\mathrm{w}(r_\mathrm{A})$}&\multicolumn{2}{c}{$T_\mathrm{p_0}$} &
 \multicolumn{2}{c}{$n_\mathrm{p_0}$}\\
 {\scriptsize [R$_{\ast}$]} &\multicolumn{2}{c}{[\scriptsize M$_{\sun}$
 yr$^{-1}$]}  &\multicolumn{2}{c}{[\scriptsize M$_{\sun}$ yr$^{-1}$]}
 &\multicolumn{2}{c}{\scriptsize [$10^6$
 cm$^{-3}$]}&\multicolumn{2}{c}{\scriptsize [K]}&
 \multicolumn{2}{c}{\scriptsize [cm$^{-3}$]}\\
 \hline
 {12} &\multicolumn{2}{c}{$2.4 ~ 10^{-10}$}  &\multicolumn{2}{c}{$1.0 ~
 10^{-11}$} &\multicolumn{2}{c}{2.91} &\multicolumn{2}{c}{$9.54 ~ 10^4$}
 &\multicolumn{2}{c}{$2.17 ~ 10^9$}  \\
 {13} &\multicolumn{2}{c}{$1.6 ~ 10^{-10}$}  &\multicolumn{2}{c}{$6.3 ~
 10^{-12}$} &\multicolumn{2}{c}{1.62} &\multicolumn{2}{c}{$6.80 ~ 10^4$}
 &\multicolumn{2}{c}{$1.73 ~ 10^9$}  \\
 {14} &\multicolumn{2}{c}{$1.0 ~ 10^{-10}$}  &\multicolumn{2}{c}{$3.6 ~
 10^{-12}$} &\multicolumn{2}{c}{0.94} &\multicolumn{2}{c}{$4.69 ~ 10^4$}
 &\multicolumn{2}{c}{$1.35 ~ 10^9$}  \\
 {15} &\multicolumn{2}{c}{$7.2 ~ 10^{-11}$}  &\multicolumn{2}{c}{$2.4 ~
 10^{-12}$} &\multicolumn{2}{c}{0.56} &\multicolumn{2}{c}{$3.55 ~ 10^4$}
 &\multicolumn{2}{c}{$1.12 ~ 10^9$}  \\
 {16} &\multicolumn{2}{c}{$5.0 ~ 10^{-11}$}  &\multicolumn{2}{c}{$1.6 ~
 10^{-12}$} &\multicolumn{2}{c}{0.35} &\multicolumn{2}{c}{$2.64 ~ 10^4$}
 &\multicolumn{2}{c}{$9.18 ~ 10^8$}   \\
 {17} &\multicolumn{2}{c}{$3.6 ~ 10^{-11}$}  &\multicolumn{2}{c}{$1.1 ~
 10^{-12}$} &\multicolumn{2}{c}{0.22} &\multicolumn{2}{c}{$2.02 ~ 10^4$}
 &\multicolumn{2}{c}{$7.66 ~ 10^8$}  \\
 \hline
 
 \end{tabular}
 \end{center}
 \end{table}
~\\

We rejected all the solutions that do not satisfy
the condition of equality between 
the thermal plasma energy density and the wind ram pressure 
in the equatorial region of the Alfv\'en surface.

As previously discussed,
the filling of the inner magnetosphere by thermal plasma is a consequence
of the radiatively driven stellar wind, which accumulates material until a steady state, 
along lines of force of the magnetic field, is established, i.e.  
the thermal plasma pressure $\epsilon_\mathrm{p}$
balances the wind ram pressure $p_\mathrm{ram}$.
This condition at the Alfv\'en point can be written as:
\begin{displaymath}
\epsilon_\mathrm{p}=p_\mathrm{ram}
\end{displaymath}
or
\begin{equation}
k_\mathrm{B} n_\mathrm{p}(r_\mathrm{A}) T_\mathrm{p}(r_\mathrm{A}) =
m_\mathrm{H} n_\mathrm{w}(r_\mathrm{A}) v_{\infty} ^2.
\label{ram}
\end{equation}
The stellar rotation implies
$n_\mathrm{p}=n_\mathrm{p_0} r^{-1}$ and $T_\mathrm{p}=T_\mathrm{p_0} r$,
hence $\epsilon_\mathrm{p}$ is 
constant inside the inner magnetosphere,
whereas the $p_\mathrm{ram}$ is defined by the
wind velocity $v_{\infty}$ and wind density 
at $r_\mathrm{A}$ (Babel \& Montermle \cite{bm97}).
Combining Eq.~\ref{ram} and Eq.~\ref{fit}, we can derive the
temperature and density of the thermal plasma trapped in the inner magnetosphere. 
The so-derived values of $T_\mathrm{p_0}$ and $n_\mathrm{p_0}$
are reported in Table~\ref{massloss} (last two columns).

\begin{figure}
\resizebox{9cm}{!}{\includegraphics{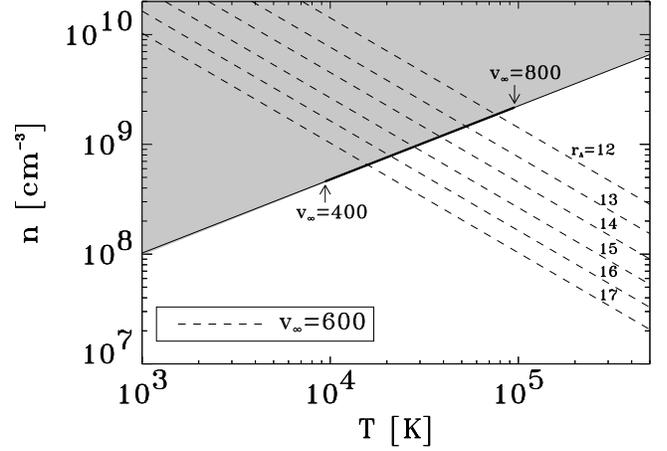}}
\caption[ ]{$\log T - \log n$ plane for the magnetospheric plasma.
In the shaded area the cold torus has $\tau_{\rm 15GHz} > 1$.
For the inner magnetosphere, the thermal pressure along the dashed lines 
equals the wind ram pressure, as computed for the possible values of
$r_\mathrm{A}$.
The thick segment of the continuous line (Eq.~\ref{fit})
represents the loci where the $T_\mathrm{p_0}$ and
$n_\mathrm{p_0}$ of our model reproduce the observations.}
\label{n_t}
\end{figure}

In Fig.~\ref{n_t}, the shaded area represents the pairs of $\log T$ and $\log n$
for which the thermal plasma is optically thick
at $\nu=$15 GHz (cold torus), while 
the Eq.~\ref{fit} and Eq.~\ref{ram}, for different values of $r_\mathrm{A}$, are 
represented by the continuous and dashed line respectively.
The intersection, marked with a thick segment, defines the ranges of values 
of $T_\mathrm{p_0}$ and $n_\mathrm{p_0}$ 
that satisfy the equality condition 
and correctly reproduce the observed radio light curves.

\subsection{Dependence on the stellar parameters}
\label{dep_stellar_param}

Since the distance and stellar radius of Cu~Vir are known with
good accuracy, the major uncertainty on the assumed parameters is
related to the magnetic field strength and geometry.
In particular, the value of $B_\mathrm{p}$ reported in the literature
ranges from 2200 to 3800 Gauss (Landstreet et al. 1977; TLL00). 
We thus examined whether our choice of $B_\mathrm{p}$=3000 Gauss
significantly influences the model results. 
We verified that a variation of $\pm$25\% in $B_\mathrm{p}$ results in
a variation of $\mp$25\% in $n_\mathrm{r}$, while the other parameters
remain unaffected.

%

\begin{figure}
\resizebox{11cm}{!}{\includegraphics{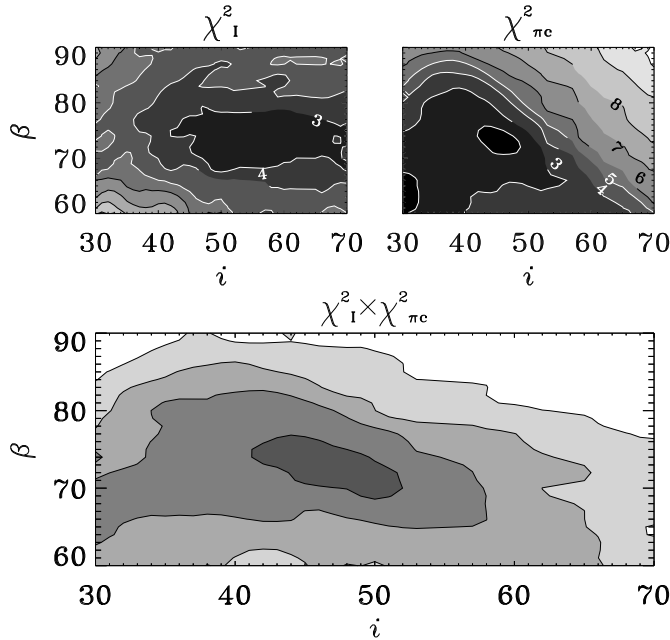}}
 \caption[ ]{Top panels: Maps, as a functions of $i$ and $\beta$, of the normalized $\chi^2$ obtained for 
the light curves at 8.4 GHz,
total flux (left) and fraction of circular polarization (right).
The $\chi^2$ minima are respectively $\approx 2$ and $\approx 1.6$.
Bottom panel: product of the two top panels.}
\label{ibeta}
\end{figure}

Moreover, we tested the sensitivity of the model to values of
inclination of the rotation axis ($i$) and obliquity of the magnetic axis ($\beta$)
different from those adopted in our calculations.
We minimize the normalized $\chi^2$ function
for the light curves (total flux density and the fraction of circular 
polarization) at 8.4 GHz.
To restrict the computational time, we limit this analysis 
to the set of model parameters associated with the lowest value of
the Alfv\'en radius ($r_\mathrm{A}=12$ R$_{\ast}$).
The maps of $\chi^2$
are displayed in Fig.~\ref{ibeta} (top panels). 
Dark areas correspond to pairs of $i$ and $\beta$ with lower $\chi^2$,
for which we find the best match between simulations and observations.
We note that the $\chi^2$ from the total flux fit is 
not well correlated to the $\chi^2$ relative to the fit of the circular 
polarization. 
The product of these two matrices, Fig.~\ref{ibeta} (bottom panel), gives the
combination of $i$ and $\beta$ that better reproduces both total flux and circular 
polarization.
We note that the most probable values of $i$ and $\beta$ are restricted in a well defined 
range of values, respectively:
$i=41{\degr} - 52 {\degr}$ and $\beta= 68{\degr} - 77 {\degr}$.
We note that the values of $i$ and $\beta$ from the literature are well inside the previous intervals.

\begin{figure*}
\resizebox{16cm}{!}{\includegraphics{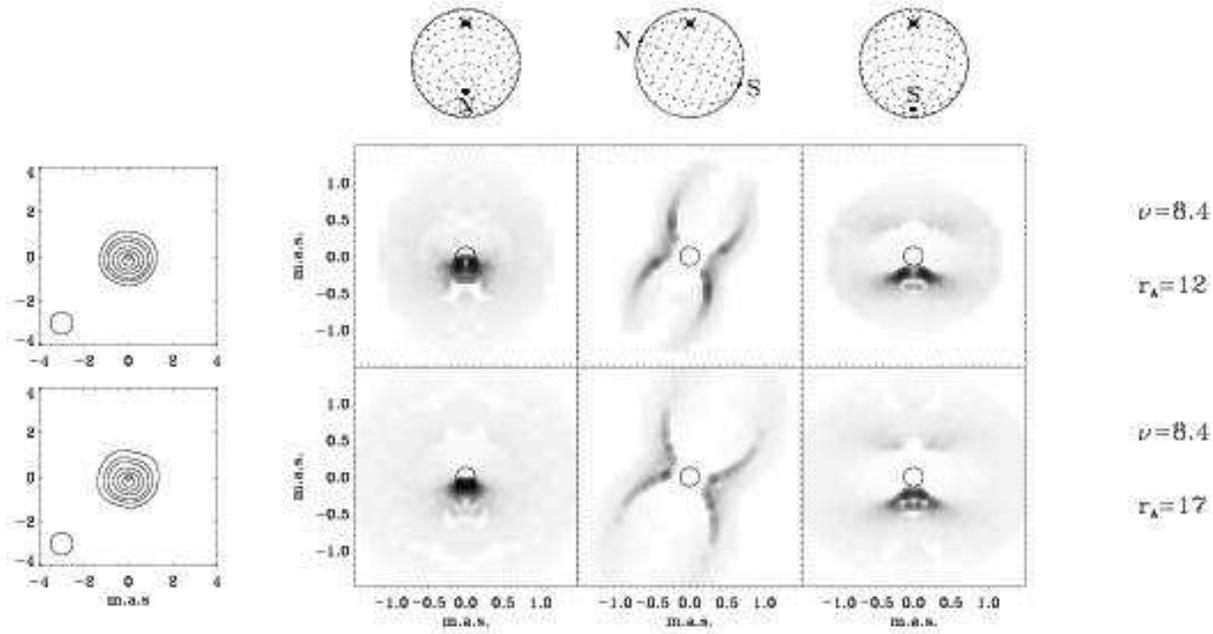}}
 \caption[ ]{Simulated maps for stellar orientations associated with the
radio light curve extrema.
On the left side the corresponding radio
maps are displayed; the theoretical bightness
spatial distribution is averaged over half the
rotational period and convolved  with the 1 mas
beam of the VLBI array.
The contours are $10\%$, $20\%$, $30\%$, $50\%$, $70\%$, $90\%$ of the peak brightness.}
\label{film}
\end{figure*}

\subsection{Limits of the model} 

While the modulation of the total flux density light curve observed
at 5 GHz is well reproduced by our simulations, we are 
not able to match the two peaks observed
close to the phases 0.40 and 0.85 of the circularly polarized flux light curve.
These are also the phases of the
coherent emission observed at 1.4 GHz (Fig.~\ref{dat_dat}, bottom panel), 
a coincidence suggesting that the two peaks are residuals
of the Electron Cyclotron Maser Emission (ECME) mechanism (TLL00).
As the ECME mechanism is not considered in our model,
which computes only incoherent gyrosynchrotron, it cannot
reproduce these observed peaks.

At 15 GHz,
a stronger and broader polarization is expected in the range
$\Phi =$ 0.9 -- 1.3 (Fig.~\ref{oss-tot}). This discrepancy could be ascribed to the simple dipolar configuration
adopted for the magnetic field. In the magnetospheric regions close
to the star, where the bulk of 15 GHz radiation is emitted, the high order
components of the magnetic field of CU~Vir (Hatzes \cite{h97}; Kushinig et al. \cite{k_etal99}; TLL00;
Glagolevskij \& Gerth \cite{gg02}) cannot be neglected.
A simple dipole provides
a symmetric total flux light curve and an enhancement of the fraction of polarization 
when the line of sight is
close to the magnetic pole, where the magnetic lines 
are almost parallel. In the presence of multipolar components
the radio flux light curves would not be symmetric
and the fraction of the polarization would be reduced.

\section{Discussion}

In the following we will analyze physical information
about the stellar magnetosphere that can be inferred from the results of our
simulations.


\subsection{The acceleration process}
The convergence of the value of the spectral index of the emitting electron distribution
($N(E)\propto E^{-\delta}$) toward $\delta=2$ 
indicates that the acceleration
process occurs in the current sheets just outside the Alfv\'en radius.
This value agrees with the one
derived in the case of solar flares by Zharkova \& Gordovskyy (\cite{zg05}), who
found a spectral index $\delta=2$ for the non-thermal electrons accelerated in the neutral
reconnecting current sheets of solar magnetic loops.
In addition, from the ratio between $n_\mathrm{r}$ (Table~\ref{sol_par}) and the density of the electrons of the wind at 
the Alfv\'en point $n_\mathrm{w}(r_\mathrm{A})$ (Table~\ref{massloss}), 
we can estimate that the efficiency of the acceleration process
for non-thermal electrons is 
$2\times 10^{-3\pm1}$. The efficency inferred by our analysis
is consistent with the
acceleration efficiency in large stationary unstable current sheets associated
with coronal loops during solar flares (Vlahos et al. \cite{v_etal04}).

\subsection{The radio emitting region}

We investigated the possibility 
of discrimining among the previous possible values of the Alfv\'en radius by using high angular resolution
radio observations in the future.
In Fig.~\ref{film} the brightness spatial distributions at 8.4 GHz obtained for the 
extreme values of $r_\mathrm{A}$ (respectively 12 and 17 $R_{\ast}$) are shown.
The simulated radio maps have been calculated for three rotational phases: 
the stellar orientations that show the north and south poles, associated with the two maxima and the
orientation that gives a null effective magnetic field, associated with the minimum of
the radio light curve.
In the right panel of Fig.~\ref{film}, the different extension of the  
radio source for the two cases analyzed is clearly visible.
However such a difference is not so obvious when looking at the radio emission arising from regions 
near the stellar surface, where the highest fraction of the magnetospheric radio 
emission originates.

For the Very Long Baseline Interferometry (VLBI)
technique, that can be used  
to map CU~Vir at milli arcsecond scales, we derived two ``synthetic'' maps at VLBI resolution,
shown on the left side of Fig.~\ref{film}, 
obtained as the average of hypothetical 8 hours observation around the stellar phase coinciding with the null
effective magnetic field, convolved with a Gaussian 1 mas beam.
It  is clear that the present capability of VLBI observations 
does not allow us to discriminate between the extreme values of the Alfv\'en
radius of CU~Vir predicted by our model.

\subsection{The thermal plasma}
The physical condition in the inner magnetosphere play an important role in the emerging
radio flux because of the free-free absorption by the plasma, which affects the modulation
of the radio light curves and the spectrum of the source.
The comparison between our model prediction and multifrequency observations
reveals the presence of a torus of material colder and denser than the trapped thermal
plasma. This is necessary to reproduce the correct scale of the total flux density in
the 15 GHz light curve of CU~Vir. 
The electron number density ($n_\mathrm{ct}$) and temperature ($T_\mathrm{ct}$)
of the plasma of the cold torus, are within the shaded area
of the plane $\log T-\log n$ shown in Fig.~\ref{n_t}, 
where the optical depth of the cold torus is higher than 1 at 15 GHz.
Higher frequency measurements (for example 22 and 43 GHz)
could characterize the spatial distribution of the thermal plasma in the deep magnetospheric regions,
near the stellar surface.

\begin{figure}
\resizebox{9cm}{!}{\includegraphics{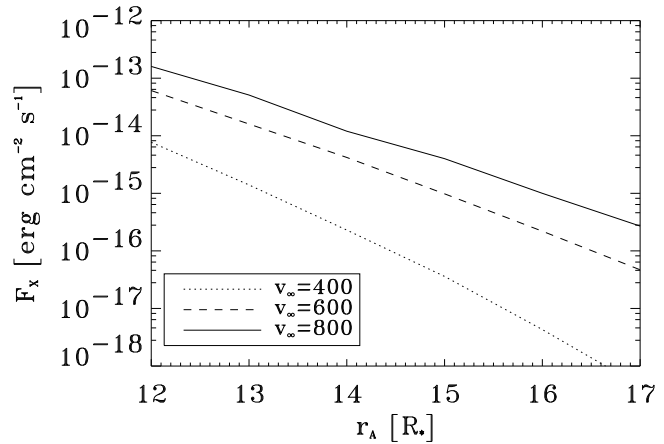}}
\caption[ ]{Predicted X-ray fluxes of the inner magnetosphere of CU~Vir
for the solutions of our 3-D model.
}
\label{LX}
\end{figure}

The temperature and density of the thermal plasma trapped in the inner magnetosphere 
are assigned for a given Alfv\'en radius and wind velocity (Table~\ref{massloss}).
As a by-product of our analysis, we estimate the X-ray emission from the thermal 
plasma trapped in the stellar inner magnetosphere of
CU~Vir as a function of $r_\mathrm{A}$ and $v_{\infty}$.
Following the same procedure described in TLU04, 
for each element of the 3D cubic grid that samples
the magnetosphere, we computed the thermal bremsstrahlung emission coefficients and
the emerging power integrated between 0.1 and 10 keV,
in the typical energy range of X-ray telescopes like Chandra and XMM.
The resulting X-ray fluxes at the Earth
are displayed in Fig.~\ref{LX}.

The stellar X-ray emission is a function of the size of the Alfv\'en surface (Fig.~\ref{LX})
and, in particular, it decreases as 
the Alfv\'en radius increases. This is a consequence of the
lower temperature and density of the plasma in the inner magnetosphere 
for increasing $r_\mathrm{A}$.
For the same reason, lower wind velocities cause a lower X-ray emission.

On the basis of the values derived from our calculations, we conclude that the
X-ray emission from CU~Vir is below the detection limit of
the ROSAT all sky survey (the average background at the position
of CU~Vir in the RASS is higher than $10^{-12}$ erg cm$^{-2}$ s$^{-1}$).
Nevertheless, for exposure times of a few tens of kilo-seconds,
Chandra and XMM can reach an 
X-ray detection limit close to  $10^{-15}$ erg cm$^{-2}$ s$^{-1}$,
which is lower than the X-ray fluxes predicted from several
model solutions. 

Within our model,
measurement of the X-ray flux of CU~Vir could place an important
constraint on the physical condition in the magnetosphere.

\section{Conclusion}
The radio magnetosphere of CU~Vir has been extensively studied
using of the first multifrequency VLA observations, covering the
entire rotational period, for both the total flux density (Stokes\,$I$)
and the circular polarization (Stokes\,$V$). 
We have successfully modeled the observed flux modulation and the
circular polarization light curves using the 3D model developed by us (TLU04),
modified to resolve the circular polarization intensity.

We pointed out the importance of multi-frequency and dual polarization
observations throughout the rotational period to gain insight into the physical conditions of the
circumstellar environments surrounding MCP stars.

Our results are:
\begin{itemize}
\item[-] We confirm the physical scenario proposed by Andr\'e et al. (\cite{a_etal88}) and
Linsky et al. (\cite{l_etal92}) and the MCWS model proposed by Babel \& Montermle (\cite{bm97});

\item[-] The size of the CU~Vir inner magnetosphere has been constrained in the
range: $r_\mathrm{A} = 12 - 17$ R$_{\ast}$;

\item[-] we evaluated the mass loss rate of CU~Vir, which up to now was not determined for this 
star, to be about $10^{-12}$ M$_{\sun}$ yr$^{-1}$;

\item[-] the temperature of the thermal electrons trapped in the inner magnetosphere 
can reach $10^5-10^6$~K and can give detectable X-ray emission;

\item[-] the presence of an accumulation of cold and dense material around
the star has been shown and limits for its size and physical conditions
(temperature and density) are given; 

\item[-] the acceleration in the current sheets produces a non-thermal electron
population with a hard energetic spectrum ($N(E)\propto E^{-\delta}$ with $\delta=2$),
and the acceleration process has an efficiency of about $10^{-3}$.

\end{itemize}

The limits of our model to reproduce all the observed features,
for example the two peaks at phases $\approx 0.4$ and $\approx 0.85$ 
observed in the 5 GHz light curve for the
circular polarization and the shape of the light curves at 15 GHz,  
have been interpreted respectively as due to the residual of the ECME
mechanism (responsible for the features previously observed at 1.4 GHz, TLL00)
and as evidence of a multipolar component in the magnetic field of CU~Vir.

\begin{acknowledgement}
We thank the referee for his/her constructive criticism which
enabled us to improve this paper.
\end{acknowledgement}%

\appendix
\section{Calculation of circular polarization}
\label{app}

The circularly polarized intensity may be obtained using
the relation, given by Ramaty (\cite{r69}):
\begin{equation}
V=2\left( \frac{I_{+} {a_{\theta}}_+}{1+{{a_{\theta}}_+}^2}+\frac{I_{-} {a_{\theta}}_-}{1+{{a_{\theta}}_-}^2} \right)
\label{v_pol1}
\end{equation}
where $I_{\pm}$ are solutions of the radiative transfer
equation for the ordinary (sign ``$+$'') and
extraordinary (sign ``$-$'') modes (Klein \& Trotter \cite{kt84}):
\begin{equation}
\frac{d}{ds} \frac{I_{\pm}|\cos{\psi_{\pm}}|}{{{n_{\theta}}_{\pm}}^2}=\frac{|\cos{\psi_{\pm}}|}{{{n_{\theta}}_{\pm}}^2} (\eta_{\pm}-k_{\pm}I_{\pm})
\label{v_pol}
\end{equation}
with ${\eta}_{\pm}$ and $k_{\pm}$ gyrosynchrotron emission
and absorption coefficients,
${n_{\theta}}_{\pm}$ and ${a_{\theta}}_{\pm}$ 
refractive index and polarization coefficient for the two modes respectively. $\psi_{\pm}$
is the angle between the wave vector and the group velocity,
given by the equation:
\begin{displaymath}
\psi_{\pm}=\arctan \left( \frac{\partial {n_{\theta}}_{\pm} }
{{n_{\theta}}_{\pm}\partial\theta} \right)
\end{displaymath}
with $\theta$ the angle between the wave vector and the magnetic field.
The equation~\ref{v_pol} was numerically solved,
along a path parallel to the line of sight,
using the method developed by us in TLU04.

In accordance with classical physics,
for propagation parallel to the magnetic field lines ($\theta=0$
and ${a_{\theta}}_{\pm} = \pm 1$),
the ordinary mode is characterized by the electric field vector rotating
in the sense of the electrons, counterclockwise in the case of a wave approaching the observer,
therefore the polarization sense for the ordinary mode is LCP; conversely
the extraordinary mode has a polarization sense RCP.

VLA measurements of the wave circular polarization state are in accordance
with the IAU and IEEE orientation/sign convention, unlike the classical physics usage.
Therefore to compare our simulations with the VLA observations 
we have to change
the sign of the result of equation~\ref{v_pol1}.

\end{document}

%% file: tab5_cuvir.tex

\footnotesize
\begin{center}
\begin{tabular}{cccrcccrccc}
\hline
\hline
 {\footnotesize $\langle$UT$\rangle$} & {\footnotesize $I_5~~(\sigma)$} 	&{\footnotesize $V_5~~(\sigma)$}	&{~~~~}
& {\footnotesize $\langle$UT$\rangle$} & {\footnotesize $I_{8.4}~~(\sigma)$}	&{\footnotesize $V_{8.4}~~(\sigma)$}	&{~~~~}
& {\footnotesize $\langle$UT$\rangle$} & {\footnotesize $I_{15}~~(\sigma)$}	&{\footnotesize $V_{15}~~(\sigma)$}	\\
 &{\scriptsize [mJy]} &{\scriptsize [mJy]} &{~~~~}
& &{\scriptsize [mJy]} &{\scriptsize [mJy]} &{~~~~}
& &{\scriptsize [mJy]} &{\scriptsize [mJy]} \\
\hline
\multicolumn{11}{c}{\scriptsize Date: 1998 June 02}\\
\hline
 {\scriptsize 00:30:45}  &{\scriptsize $2.78~~(0.05)$} &{\scriptsize $<3\sigma~~(0.05)$} &{~~~~}
& {\scriptsize 01:37:00}  &{\scriptsize $3.18~~(0.04)$} &{\scriptsize $<3\sigma~~(0.04)$} &{~~~~}
& {\scriptsize 02:51:50}  &{\scriptsize $4.1~~(0.1)$} &{\scriptsize $<3\sigma~~(0.1)$} 	\\
 {\scriptsize 01:23:05}  &{\scriptsize $3.27~~(0.05)$} &{\scriptsize $~~0.32~(0.05)$} &{~~~~}
& {\scriptsize 02:36:55}  &{\scriptsize $3.87~~(0.04)$} &{\scriptsize $<3\sigma~~(0.04)$} &{~~~~}
& {\scriptsize 03:51:35}  &{\scriptsize $4.3~~(0.1)$} &{\scriptsize $<3\sigma~~(0.1)$} 	\\
 {\scriptsize 02:22:55}  &{\scriptsize $3.64~~(0.04)$} &{\scriptsize $<3\sigma~~(0.04)$} &{~~~~}
& {\scriptsize 03:38:50}  &{\scriptsize $4.40~~(0.04)$} &{\scriptsize $<3\sigma~~(0.04)$} &{~~~~}
& {\scriptsize 04:51:25}  &{\scriptsize $3.4~~(0.1)$} &{\scriptsize $<3\sigma~~(0.1)$} 	\\
 {\scriptsize 03:22:45}  &{\scriptsize $4.13~~(0.04)$} &{\scriptsize $<3\sigma~~(0.04)$} &{~~~~}
& {\scriptsize 04:36:35}  &{\scriptsize $4.42~~(0.03)$} &{\scriptsize $<3\sigma~~(0.03)$} &{~~~~}
& {\scriptsize 05:51:20}  &{\scriptsize $3.3~~(0.1)$} &{\scriptsize $<3\sigma~~(0.1)$} 	\\
 {\scriptsize 04:22:35}  &{\scriptsize $4.44~~(0.04)$} &{\scriptsize $<3\sigma~~(0.04)$} &{~~~~}
& {\scriptsize 05:36:30}  &{\scriptsize $3.69~~(0.03)$} &{\scriptsize $<3\sigma~~(0.03)$} &{~~~~}
& {\scriptsize 06:51:15}  &{\scriptsize $2.1~~(0.1)$} &{\scriptsize $<3\sigma~~(0.1)$} 	\\
 {\scriptsize 05:22:25}  &{\scriptsize $3.48~~(0.04)$} &{\scriptsize $<3\sigma~~(0.04)$} &{~~~~}
& {\scriptsize 06:36:20}  &{\scriptsize $2.85~~(0.03)$} &{\scriptsize $<3\sigma~~(0.03)$} &{~~~~}
& {\scriptsize 07:44:20}  &{\scriptsize $3.0~~(0.1)$} &{\scriptsize $<3\sigma~~(0.1)$} 	\\
 {\scriptsize 06:22:15}  &{\scriptsize $2.78~~(0.04)$} &{\scriptsize $<3\sigma~~(0.04)$} &{~~~~}
& {\scriptsize 07:30:25}  &{\scriptsize $2.85~~(0.04)$} &{\scriptsize $~~0.22~(0.04)$} &{~~~~}
& {\scriptsize \dots}  &{\scriptsize \dots} &{\scriptsize \dots} 	\\
 {\scriptsize 07:18:55}  &{\scriptsize $3.96~~(0.05)$} &{\scriptsize $~~0.58~(0.05)$} &{~~~~}
& {\scriptsize \dots}  &{\scriptsize \dots} &{\scriptsize \dots} &{~~~~}
& {\scriptsize \dots}  &{\scriptsize \dots} &{\scriptsize \dots} 	\\
 {\scriptsize 08:13:45}  &{\scriptsize $3.57~~(0.05)$} &{\scriptsize $~~0.27~(0.05)$} &{~~~~}
& {\scriptsize \dots}  &{\scriptsize \dots} &{\scriptsize \dots} &{~~~~}
& {\scriptsize \dots}  &{\scriptsize \dots} &{\scriptsize \dots} 	\\
\hline
\multicolumn{11}{c}{\scriptsize Date: 1998 June 07}\\
\hline
 {\scriptsize 02:10:15}  &{\scriptsize $4.87~~(0.04)$} &{\scriptsize $~~0.43~(0.04)$} &{~~~~}
& {\scriptsize 02:23:40}  &{\scriptsize $4.85~~(0.03)$} &{\scriptsize $~~0.77~(0.03)$} &{~~~~}
& {\scriptsize 02:38:10}  &{\scriptsize $3.3~~(0.1)$} &{\scriptsize $~0.4~~(0.1)$} 	\\
 {\scriptsize 03:08:35}  &{\scriptsize $4.87~~(0.04)$} &{\scriptsize $~~0.37~(0.04)$} &{~~~~}
& {\scriptsize 03:21:35}  &{\scriptsize $4.97~~(0.03)$} &{\scriptsize $~~0.56~(0.03)$} &{~~~~}
& {\scriptsize 03:35:55}  &{\scriptsize $3.0~~(0.1)$} &{\scriptsize $<3\sigma~(0.1)~$} 	\\
 {\scriptsize 04:06:25}  &{\scriptsize $4.68~~(0.04)$} &{\scriptsize $~~0.34~(0.04)$} &{~~~~}
& {\scriptsize 04:19:25}  &{\scriptsize $4.52~~(0.03)$} &{\scriptsize $~~0.57~(0.03)$} &{~~~~}
& {\scriptsize 04:33:45}  &{\scriptsize $3.6~~(0.15)$} &{\scriptsize $<3\sigma~(0.15)$} 	\\
 {\scriptsize 05:04:50}  &{\scriptsize $3.73~~(0.04)$} &{\scriptsize $<3\sigma~~(0.04)$} &{~~~~}
& {\scriptsize 05:18:20}  &{\scriptsize $3.50~~(0.04)$} &{\scriptsize $~~0.28~(0.04)$} 	  &{~~~~}
& {\scriptsize 05:32:45}  &{\scriptsize $3.3~~(0.15)$} &{\scriptsize $<3\sigma~(0.15)$} 	\\
 {\scriptsize 06:03:35}  &{\scriptsize $3.79~~(0.04)$} &{\scriptsize $<3\sigma~~(0.04)$} &{~~~~}
& {\scriptsize 06:17:05}  &{\scriptsize $3.88~~(0.04)$} &{\scriptsize $<3\sigma~~(0.04)$} &{~~~~}
& {\scriptsize 06:31:35}  &{\scriptsize $3.4~~(0.15)$} &{\scriptsize $<3\sigma~(0.15)$} 	\\
 {\scriptsize 07:02:30}  &{\scriptsize $3.98~~(0.04)$} &{\scriptsize $<3\sigma~~(0.04)$} &{~~~~}
& {\scriptsize 07:15:15}  &{\scriptsize $4.01~~(0.04)$} &{\scriptsize $<3\sigma~~(0.04)$} &{~~~~}
& {\scriptsize \dots}     &{\scriptsize \dots} &{\scriptsize \dots} 	\\
 {\scriptsize 07:54:30}  &{\scriptsize $4.38~~(0.05)$} &{\scriptsize $<3\sigma~~(0.05)$} &{~~~~}
& {\scriptsize \dots}	  &{\scriptsize \dots} 		&{\scriptsize \dots} 	  &{~~~~}
& {\scriptsize \dots}  	  &{\scriptsize \dots} 		&{\scriptsize \dots} 	\\
\hline
\multicolumn{11}{c}{\scriptsize Date: 1998 June 12}\\
\hline
 {\scriptsize 01:21:35}  &{\scriptsize $3.95~~(0.05)$} &{\scriptsize $<3\sigma~~(0.05)$} &{~~~~}
& {\scriptsize 01:35:05}  &{\scriptsize $4.31~~(0.04)$} &{\scriptsize $-0.14~(0.04)$} &{~~~~}
& {\scriptsize 02:47:25}  &{\scriptsize $3.3~~(0.1)$} &{\scriptsize $<3\sigma~~(0.1)$} 	\\
 {\scriptsize 02:19:55}  &{\scriptsize $4.10~~(0.04)$} &{\scriptsize $<3\sigma~~(0.04)$} &{~~~~}
& {\scriptsize 02:32:55}  &{\scriptsize $4.18~~(0.04)$} &{\scriptsize $-0.21~(0.04)$} &{~~~~}
& {\scriptsize 03:45:15}  &{\scriptsize $2.5~~(0.1)$} &{\scriptsize $<3\sigma~~(0.1)$} 	\\
 {\scriptsize 03:17:45}  &{\scriptsize $3.47~~(0.04)$} &{\scriptsize $<3\sigma~~(0.04)$} &{~~~~}
& {\scriptsize 03:30:45}  &{\scriptsize $3.54~~(0.03)$} &{\scriptsize $<3\sigma~~(0.03)$} &{~~~~}
& {\scriptsize 04:44:10}  &{\scriptsize $2.7~~(0.1)$} &{\scriptsize $<3\sigma~~(0.1)$} 	\\
 {\scriptsize 04:16:05}  &{\scriptsize $3.12~~(0.04)$} &{\scriptsize $<3\sigma~~(0.04)$} &{~~~~}
& {\scriptsize 04:29:35}  &{\scriptsize $3.05~~(0.03)$} &{\scriptsize $~~0.16~(0.03)$} &{~~~~}
& {\scriptsize 05:43:00}  &{\scriptsize $3.3~~(0.1)$} &{\scriptsize $<3\sigma~~(0.1)$} 	\\
 {\scriptsize 05:15:05}  &{\scriptsize $3.28~~(0.04)$} &{\scriptsize $~~0.50~(0.04)$} &{~~~~}
& {\scriptsize 05:28:25}  &{\scriptsize $3.58~~(0.03)$} &{\scriptsize $~~0.39~(0.03)$} 	  &{~~~~}
& {\scriptsize 06:41:50}  &{\scriptsize $3.7~~(0.1)$} &{\scriptsize $<3\sigma~~(0.1)$} 	\\
 {\scriptsize 06:13:55}  &{\scriptsize $4.38~~(0.04)$} &{\scriptsize $~~0.44~(0.04)$} &{~~~~}
& {\scriptsize 06:27:20}  &{\scriptsize $4.33~~(0.03)$} &{\scriptsize $~~0.53~(0.03)$} &{~~~~}
& {\scriptsize \dots}     &{\scriptsize \dots} &{\scriptsize \dots} 	\\
 {\scriptsize 07:09:30}  &{\scriptsize $4.69~~(0.04)$} &{\scriptsize $~~0.43~(0.04)$} &{~~~~}
& {\scriptsize 07:20:55}  &{\scriptsize $4.67~~(0.04)$} &{\scriptsize $~~0.93~(0.04)$} &{~~~~}
& {\scriptsize \dots}     &{\scriptsize \dots} &{\scriptsize \dots} 	\\
\hline
\hline
\end{tabular}
\end{center}
